# When to encourage using Gaussian regression for feature selection tasks with time-to-event outcome

Rong Lu, PhD*

*The Quantitative Sciences Unit, Division of Biomedical Informatics Research, Department of Medicine, Stanford University, Stanford, California

**Corresponding Author:** Rong Lu, PhD (https://orcid.org/0000-0003-4321-9144); ronglu@stanford.edu; 3180 Porter Drive, Palo Alto, CA 94304.

**Conflicts of Interest:** None

**Funding Sources:** This work is partially supported by the Biostatistics Shared Resource (BSR) of the NIH-funded Stanford Cancer Institute (P30CA124435) and the Stanford Center for Clinical and Translational Research and Education (UL1-TR003142).

**Role of the Funder/Sponsor:** The funders had no role in the design and conduct of the study; collection, management, analysis, and interpretation of the data; preparation, review, or approval of the manuscript; and decision to submit the manuscript for publication.

**KEY POINTS**

**Question:** Which statistical methods should be used for feature selection with respect to time-to-event outcomes if the sample size is small or some of the key covariates are not measured?

**Findings:** Univariate Cox regression is not the best-performing model for feature selection or effect size ranking if the true models are multivariate Cox regression with Gaussian covariates and half of the covariates are not measured, regardless of the correlation strength between features. The regularized Cox regression with $\lambda = \lambda_{1se}$ and the Gaussian regression of log-transformed survival time with two covariates (the event indicator plus one feature at a time) are better models for feature selection when total number of events is small/modest (<500) and the true models are multivariate Cox regression with Gaussian covariates.

**Meaning:** This study demonstrates the importance of including Gaussian regression of log-transformed survival time in survival analysis when sample size is small.

**ABSTRACT**

**IMPORTANCE:** Feature selection with respect to time-to-event outcomes is one of the fundamental problems in clinical trials and biomarker discovery studies. But it's unclear which statistical methods should be used when sample size is small or some of the key covariates are not measured.

**DESIGN:** In this simulation study, the true models are multivariate Cox proportional hazards models with 10 covariates. It's assumed that only 5 out the 10 true features are observed/measured for all model fitting, along with 5 random noise features. Each sample size scenario is explored using 10,000 simulation datasets. Eight regression models are applied to each dataset to estimate feature effects, including both regularized Gaussian regression (elastic net penalty) and regularized Cox regression (glmnet Cox).

**RESULTS:** If the covariates are highly correlated Gaussian, the Gaussian regression of log-transformed survival time with only two covariates outperforms all tested Cox regression models when total number of events <500.

**INTRODUCTION**

Feature selection with respect to time-to-event outcomes is one of the fundamental problems in clinical trials and biomarker discovery studies [1-4]. Many cancer trials use either the overall survival or the progression-free survival as the primary outcome to explore or validate the efficacy of new treatments [1-3]. To this date, the Cox proportional hazards model is still the most used method for testing the effect of intervention in randomized clinical trials [3,5]. But in biomarker discovery studies that focus on screening large number of genetic markers, the regularized Cox regression seems to be more popular in recent years [6-12]. It's unclear why regularized Cox regression only gained popularity in screening large number of biomarkers, but not in analyses of randomized trials. One potential explanation is that many researchers are choosing statistical methods based on how many features were measured/available for analysis. The appropriateness of such motivation might be questionable if one believes that we should never assume anything that's not measured must have no effect on the outcome of interest.

If we want to make inference starting from the assumption that unknown/unmeasured factors can significantly affect the outcome of interest, it will be very important to study the best way of choosing statistical methods and compare different methods' performance under the same assumption. However, this assumption of unmeasured features seems to be rarely used in developing methodologies within the framework of multivariate regression models. Most simulation studies assumed that true features used for generating survival time were all available for feature selection analysis [9-17].

Motivated by this observation, this work will explore and compare the performance of several regression models in a simulation study of time-to-event outcomes, assuming that half of the true covariates are not measured for feature selection.

In literature it also seems that Gaussian regression methods are rarely used to study time-to-event outcomes, despite their wide usage in analyzing other data types where Gaussian models' assumptions cannot all hold strictly in practice [18-21]. Many methodological variants of Cox regression are being developed actively and are specifically designed for handling application challenges such as time-varying covariates/coefficients and violation of proportional hazards assumption [22-26]. While tailoring Cox regression to different application scenarios remains a very hot research topic, not much effort seems to be devoted to compare models that were not originally proposed for survival analysis with typical survival models. In this simulation study, I am also interested in testing the performance of a few simple Gaussian models in feature selection tasks with time-to-event outcomes, by using data generated from multivariate Cox proportional hazards models.

It is also observed that many researchers prefer firstly fitting univariate Cox regression models or performing literature review to select features that might have significant association with time-to-event outcomes, before including only those selected features in fitting multivariate Cox regression to estimate effect sizes [27-31]. This strategy makes intuitive sense only if the selection performed by univariate Cox regression or literature review can significantly increase the likelihood of identifying all features with large effect sizes so that the multivariate Cox regression can be fitted with exactly correct or almost correct parametric form/portion, which however will naturally require all key features to be measured and available for analysis. But it's unclear how this strategy will behave when many true covariates are not measured. I think it would be interesting to explore whether any existing methods could outperform this strategy when many key features are not measured.

In this simulation study, total eight models' performance in feature selection tasks will be explored under the assumptions that: 1) Time-to-event values are generated from multivariate Cox proportional hazards models, with different correlation levels between features, different sample sizes, different censoring rates and different parameter values for baseline risk function. 2) Not all true features are measured for feature selection tasks. True models with 10 covariates will be used to generate time to events. But only randomly selected 5 true features will be available for feature selection analysis, along with 5 noise features. 3) Censored time will occur randomly and follow Uniform distributions.

**METHODS**

In this simulation study, the true models are always multivariate Cox proportional hazards models with 10 true features/covariates for all simulation scenarios, but with different choices of parameter values for baseline cumulative risk function, feature correlation levels, sample sizes, and censoring rates. All Gaussian models tested in this study use the log-transformed survival/censored time as the response variable and include the event indicator as one of the covariates.

Parameters for generating survival and censored time:

Each simulation dataset is created using different parameter values for baseline cumulative risk function. These parameters are randomly drawn using R function *sample(c(10:200)/5,1),* which was designed to reduce the probability of generating survival time values that are too concentrated towards zero and have extremely long right tails at the same time (line 83-84 in the [example script](example script)). The 10 true features are generated using multivariate Gaussian distribution with mean $\mu=0$ and variance $\sigma^2=1$ for all features (line 88, 89, and 98 in the scripts). The covariance values of multivariate Gaussian will differ between simulation scenarios, to create either independent true features or highly correlated true features ($\rho=0$ or 0.8). The coefficients of the 10 true features are always 1, 2, 3, …, 10 (line 92 in the scripts). Censored survival time was assumed to occur randomly and follow the uniform distribution on [0, time-to-event] (script line 121 and 123). Each sample size (n = 500, 1000, 2000, or 5000; script line 86) and censoring rate scenario (10% or 50%) is explored using 10,000 simulation datasets (script line 28).

Implementation of the assumption that not all true features are measured:

In all simulation scenarios, it's assumed that only 5 out of the 10 true features are observed/measured, along with 5 noise features that are generated independently from the true features (script line 312 and 340). The 5 observed true features are randomly picked with equal probability (script line 87 and 130). Those noise features are generated using the same multivariate Gaussian distribution that is used to generate true features (script line 99).

Models of interest:

The following regression models will be applied to every simulation dataset to screen the 10 available features (5 true features + 5 noise features) and to rank their effect sizes:

1. Univariate Cox proportional hazards model: Total 10 univariate Cox models will be fitted using each simulation dataset, one for each measured feature. The performance of univariate Cox regression will be compared with other methods in terms of both feature selection accuracy and the accuracy in ranking the effect sizes (using the coefficients' p-values to rank the effect sizes; please see script line 182-193).

2. Multivariate Cox proportional hazards model without regularization: To explore the upper bound of model performance in ranking effect sizes, I assumed that in ideal situation the field experts correctly identified all 5 true features beforehand. Therefore, only the 5 true features will be included in fitting one multivariate Cox models to rank their effect sizes for each simulation dataset (using the coefficients' p-values to rank the effect sizes; please see script line 131 and line 174). That is, the multivariate Cox model (without any regularization penalty) will not be compared with other methods in terms of feature selection sensitivity and specificity.

3. <u>Multivariate Cox regression with elastic net regularization ($\lambda = \lambda_{1se}$)</u>: A 10-fold cross-validation will be performed using R function *cv.glmnet()* to estimate $\lambda_{1se}$ [9-12], the largest $\lambda$ such that error is within 1 standard error of minimum mean cross-validated error (please see script line 312-315 and line 317). One Cox regression with elastic net regularization ($\lambda = \lambda_{1se}$) will be fitted for each simulation dataset, where all measured features and the event indicator will be included as the covariates (script line 322). The features will be selected if its coefficient is non-zero. The ranking of effect sizes will be obtained by sorting the absolute value of each coefficient (script line 331).

4. <u>Multivariate Cox regression with elastic net regularization ($\lambda = \lambda_{min}$)</u>: A 10-fold cross-validation will be performed using R function *cv.glmnet()* to estimate $\lambda_{min}$ [9-12], the value of $\lambda$ that gives minimum of the mean cross-validated error (please see script line 312-316). One Cox regression with elastic net regularization ($\lambda = \lambda_{min}$) will be fitted for each simulation dataset, where all measured features and the event indicator will be included as the covariates (script line 321). The features will be selected if its coefficient is non-zero. The ranking of effect sizes will be obtained by sorting the absolute value of each coefficient (script line 327).

5. <u>Univariate logistic regression</u>: Total 10 Logistic regression models will be fitted using each simulation dataset, one for each available feature. The response variable is the event indicator. The only 2 covariates are the feature under screening and the survival/follow-up time (script line 225-233 and line 237-245). The performance of univariate logistic regression will be compared with other methods in terms of both feature selection accuracy and the accuracy in ranking the effect sizes. The ranking of effect sizes will be obtained by sorting the coefficients' p-values (please see script line 234).

6. <u>Gaussian regression with two covariates</u>: Total 10 Gaussian regression models will be fitted using each simulation dataset, one for each measured feature. The response variable is the log-transformed survival time. The 2 covariates are the feature under screening and the event indicator (script line 248-254 and line 268-274). The performance of univariate logistic regression will be compared with other methods in terms of both feature selection accuracy and the accuracy in ranking the effect sizes. The ranking of effect sizes will be obtained by sorting the coefficients' p-values (please see script line 257).

7. <u>Multivariate Gaussian regression without regularization</u>: Using the log-transformed survival time as the response variable, one multivariate Gaussian regression without regularization will be fitted using each simulation dataset, where the event indicator and only features selected by the Gaussian model with only two covariates (model #6) will be included as the covariates (script line 284-295). These features will be ranked using their coefficients' p-values (script line 305).

8. <u>Multivariate Gaussian regression with elastic net regularization ($\lambda$=0.05)</u>: Using the log-transformed survival time as the response variable, one multivariate Gaussian regression with elastic net regularization ($\lambda$=0.05) [10] will be fitted using each simulation dataset, where all measured features and the event indicator will be included as the covariates (script line 340-345). Features will be selected if their coefficients are not equal zero. The ranking of effect sizes will be obtained by sorting the absolute value of each coefficient (script line 348).

<u>Assessment of Model Performance:</u>

Model performance will be assessed using sensitivity, specificity, and the probability of correctly ranking all true effect sizes. For the calculation of all performance statistics, each simulation dataset is considered as one analysis unit. That is, the sample size for estimating each performance statistic is always 10,000 in any simulation scenario. Although within the same simulation scenario where the sample sizes and censoring rates are equal across all 10,000 datasets, the parameters of baseline cumulative risk function are randomly selected for each dataset independently (please see variable lambda and alpha in script line 11, 25 and 83-84). That is, the 10,000 simulation datasets within each scenario cover a wide range of different shapes of baseline cumulative risk function. Therefore, the reported performance estimates are summary statistics cover a wide range of different shapes of baseline cumulative risk function as well. For the calculation of sensitivity and specificity, each analysis unit (or simulation dataset) is considered positive-by-test only if all true features are correctly identified and is considered negative-by-test only if all noise features are correctly identified. Similarly, for the calculation of accuracy of effect size ranking, each analysis unit is considered having accurate ranking only if all true features are correctly ranked.

To help clarify all methodology details and ensure reproducibility of this work, all scripts and scripts' outputs are provided in a [code repository](#) with public access. These scripts can also be modified to facilitate simulation studies of other parameter or model choices, which might be needed to better imitate the survival time distribution, feature correlation structure, sample size, and censoring rate/timing that are observed in real datasets. One additional R script is provided to generate survival time using Cox regression models with baseline risk function parameters between 0 and 4 (line 42-43 of the [clarification script](#)).  This additional script can demonstrate that when baseline risk function parameter alpha gets closer to zero, simulated survival time distribution concentrates near zero and has extremely long right tail at the same time (line 85-115 of the clarification script), which often leads to failure of partial likelihood inference even if all key covariates are measured and included in fitting Cox regression ([Supplementary Figure 2](#)).

**RESULTS**

The main results are summarized in Figure 1. Panel A and C plot feature selection accuracy vs. the total number of events in each simulation scenario. For scenarios included in Panel A, all features are independent Gaussian. Scenarios in Panel C assumed that all true features were highly correlated ($\rho=0.8$). Each dot in these figures corresponds to one simulation scenario of total number of events (i.e., total sample size × (1−censoring rate)). The probability estimates of each scenario was calculated using 10,000 simulation datasets, which cover a wide range of baseline cumulative risk functions. Similarly, Panel B and D plot the accuracy of effect size ranking vs. total number events.

When features are independent Gaussian

Figure 1 Panel A indicates that Cox regression model with elastic net regularization (*glmnet* Cox model fitted with $\lambda_{1se}$) is the best method to use, in terms of achieving the highest feature selection accuracy when all features are independent. But when total number of events is small or modest (<500), multivariate Gaussian regression with elastic net regularization ($\lambda=0.05$), fitted with the event indicator and all true/noise features, outperformed all other methods by achieving the highest probability of correctly rank all true effect sizes (Figure 1 Panel B).

It's also worth noting that Gaussian regression (of log-transformed survival time with two covariates: one feature plus the event indicator) consistently outperformed both the univariate Cox proportional hazards model and the logistic regression, in terms of not only higher sensitivity, comparable specificity, but also higher accuracy of effect size ranking, regardless of the sample size and censoring rate choices (Supplementary Table 1). This is the same observation reported in another simulation study [32], where the follow-up time was simulated differently by assuming that dropouts occurred randomly but not uniformly.

Based on these observations, I believe that both the univariate Cox proportional hazards model and the logistic regression should not be recommended for screening features or estimating effect sizes when the true model is believed to be a multivariate Cox regression with independent Gaussian features. Instead, the following analysis steps will likely work better if total number of events is small/modest (<500):

- Step 1: Fit one multivariate Gaussian regression with elastic net regularization ($\lambda$=0.05) to estimate the effect sizes, using the log-transformed survival/follow-up time as the response variable, including all measured true/noise features and the event indicator as the covariates.
- Step 2: Fit one *glmnet* Cox model with all features using $\lambda_{1se}$, and then select features if the *glmnet* Cox coefficients not equal zero.
- Step 3: rank all selected features (from Step 2) using the feature coefficients reported by the regularized Gaussian model in Step 1. This ranking can be used as the main result of effect size ranking.

When features are highly correlated Gaussian

Figure 1 Panel C suggests that when all features are highly correlated and the total number of events is small/modest (<900), Gaussian regression with two covariates (the event indicator plus one feature) is the best method for achieving the highest feature selection accuracy = (sensitivity + specificity)/2. This simple method reported >90% selection accuracy even when the total number of events is as small as 50. But its effect-size-ranking accuracy is far from acceptable, especially when the sample size is large (Figure 1 Panel D). From Panel C, we can also see that the Gaussian model with elastic net regularization ($\lambda$=0.05) reported the highest selection accuracy when the total number events is large (> 900). Figure 1 Panel D shows that Cox regression models with elastic net regularization (fitted with $\lambda_{1se}$ or $\lambda_{min}$) consistently outperformed all other methods in the task of ranking effect sizes.

Similar to the results reported by a slightly different simulation design [32], the univariate Cox model reported lower feature selection accuracies than the Gaussian models and was observed with one of the worst accuracies in ranking effect sizes, especially when features are highly correlated and the number of events is large (>1000). The logistic regression reported the worst sensitivity and the worst accuracy of effect size ranking when true features are highly correlated (Supplementary Table 1).

Based on these observations, I believe that both the univariate Cox proportional hazards model and the logistic regression are again not suited for screening features or estimating effect sizes when the true model is believed to be a multivariate Cox proportional hazards model with highly correlated Gaussian features. The Gaussian models are no longer the best choices for ranking effect sizes when features are highly correlated. The following steps might work better in practice if features are highly correlated:

- Step 1: Fit one multivariate Cox regression with elastic net regularization ($\lambda = \lambda_{1se}$) to estimate the effect sizes of all features.
- Step 2: When the total number of events is small/modest (<900), multiple Gaussian regression models need to be fitted to test one feature at a time, using the log-transformed survival/follow-up time as the response variable. Every Gaussian model contains two covariates only: the event indicator and the feature under screening. Then the feature is selected if its coefficient's p-value < 0.05. But when the total number of events is large (≥900), one multivariate Gaussian regression with elastic net regularization can be fitted with $\lambda = 0.05$, using the log-transformed survival/follow-up time as the response variable, but including all measured features along with the event indicator as the covariates. Then one feature is selected if its coefficient (reported by the regularized Gaussian model) is not zero.
- Step 3: rank all selected features (from Step 2) using the feature coefficients reported by the regularized Cox model in Step 1. This ranking can be used as the main result of effect size ranking.

Other Observations of Method Sensitivity and Specificity

From Supplementary Figure 1 Panel A we can see that when features are independent and the total number of events is small or modest (<1000), both the Gaussian regression with elastic net regularization ($\lambda$=0.05) and the Cox regression with elastic net regularization ($\lambda = \lambda_{min}$) were able to achieve much higher sensitivity in feature selection tasks, compared to other methods. But both models also reported much lower specificity estimates than other methods in these simulation scenarios (Sup Figure 1 Panel B and Sup Table 1). It's also worth noting that when features are independent, the Cox regression with elastic net regularization ($\lambda = \lambda_{min}$) is the only method that reported decreasing trend in specificity (<0.2) over the increased number of events on interval [50, 4500]. And it's also interesting to see that all three "one-feature-at-a-time" methods (Univariate Cox model, Univariate Logistic regression, and Gaussian model with two covariates) reported fairly constant specificity (around 0.75-0.79) over the increased number of events on the interval [50, 4500].

When the true model is a multivariate Cox regression with highly correlated features, almost all explored methods were able to achieve perfect or nearly perfect sensitivity using small/modest number of events, except the univariate logistic regression (Sup Figure 1 Panel C). As shown in Sup Figure 1 Panel D and Sup Table 1, all three regression modes with elastic net regularization (Gaussian model with $\lambda$=0.05 and Cox model with $\lambda_{min}$ or $\lambda_{1se}$) reported much lower specificity compared to the three "one-feature-at-a-time" methods, when the total number of events is small or modest (<900). The Cox regression with elastic net regularization ($\lambda = \lambda_{min}$) reported the worst specificity (<0.2) over increased number of events on the interval [50, 4500].

[Figure 1](): Line graph of model performance estimates over total number of events

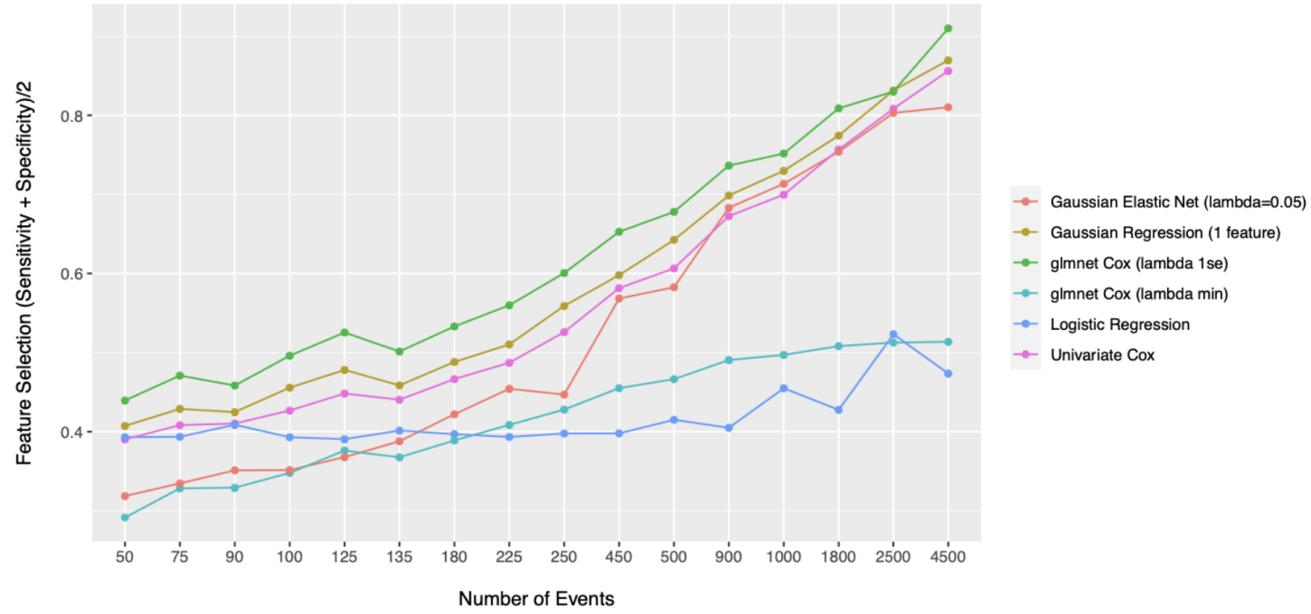

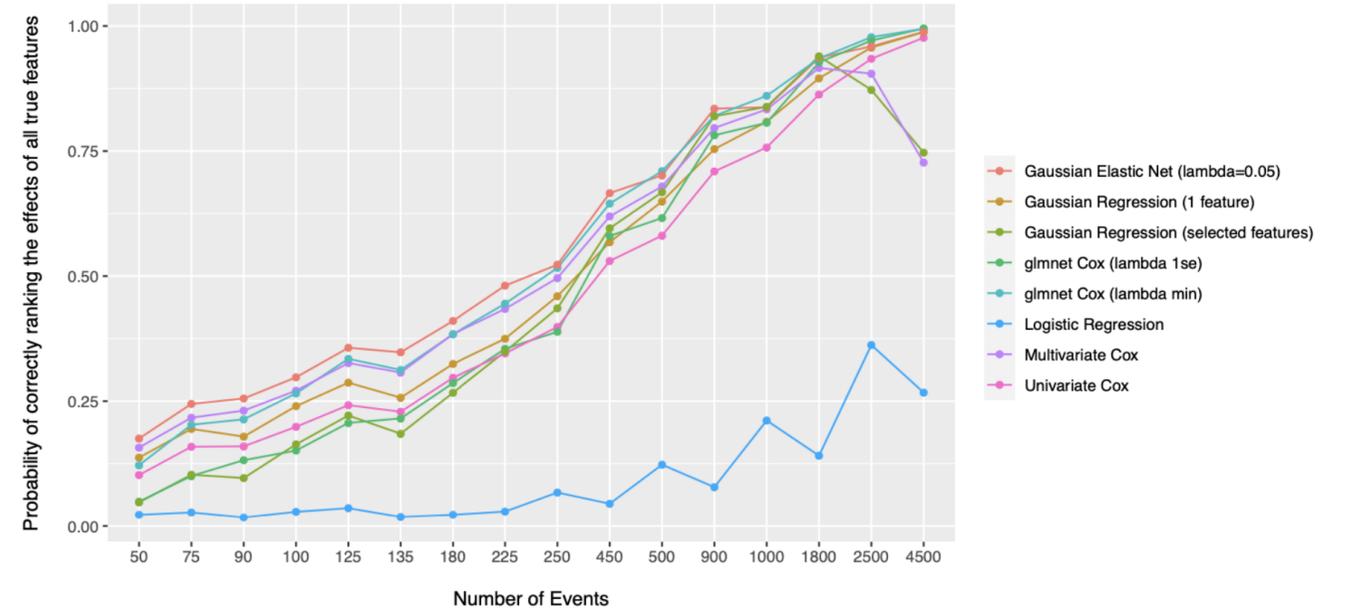

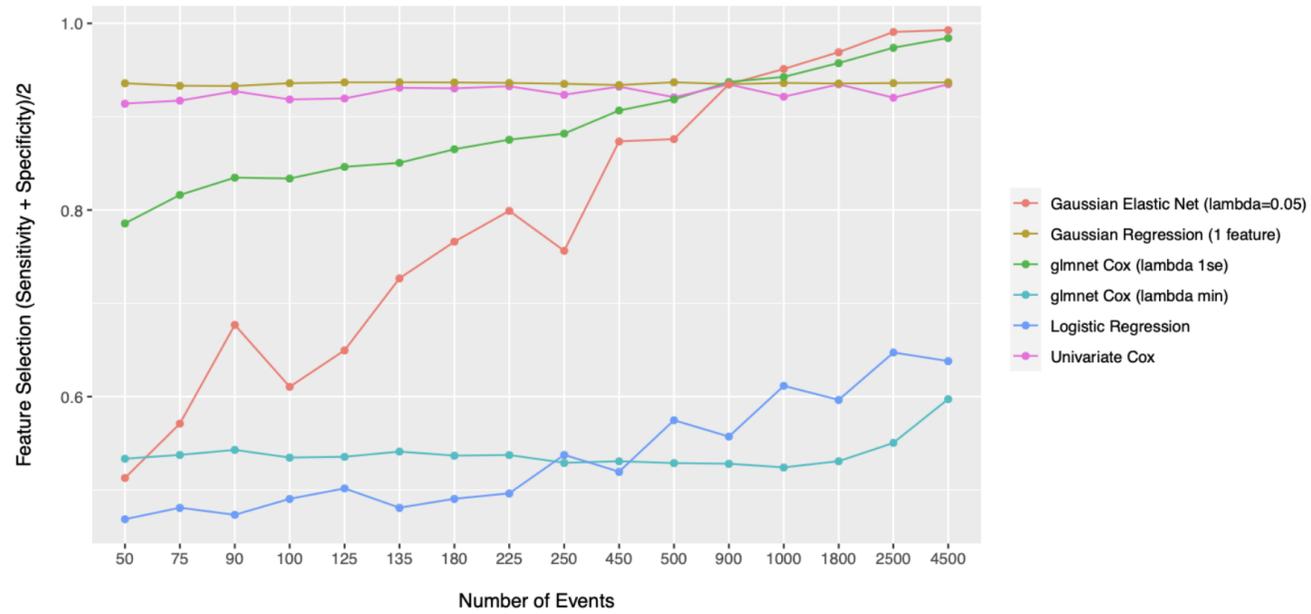

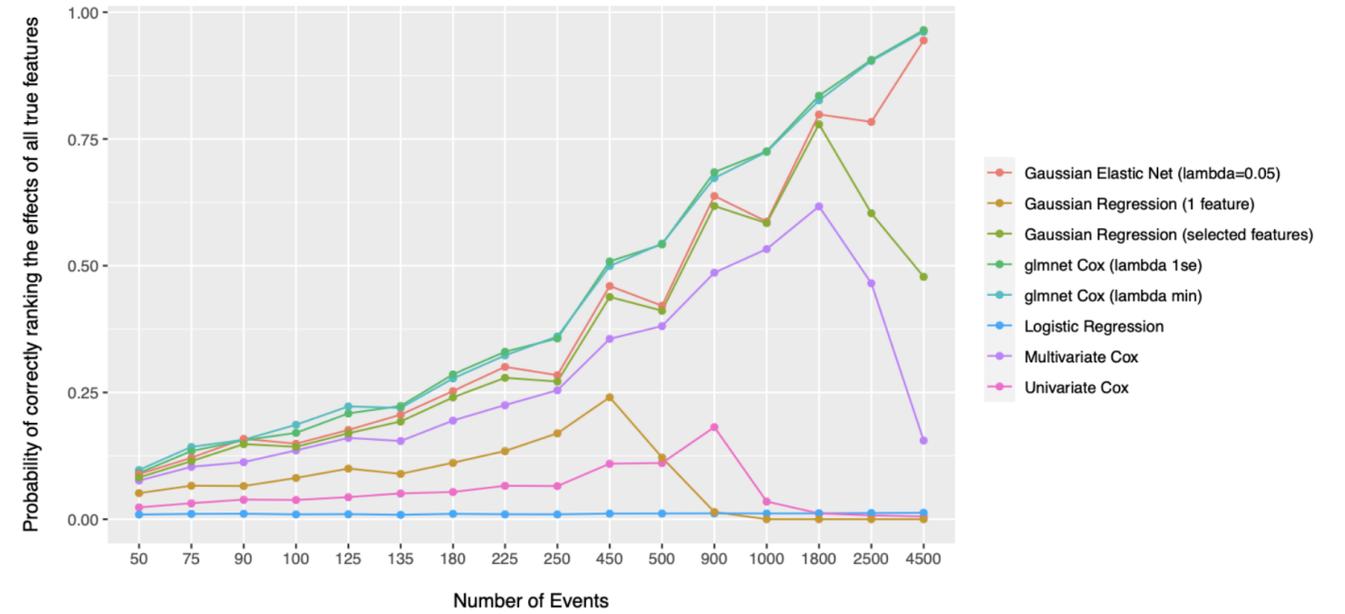

**DISCUSSION**

This simulation study presents a set of realistic scenarios where Gaussian regression of log-transformed survival time can outperform a few commonly used Cox regression methods in survival analyses, either in terms of better feature selection accuracy (Figure 1 Panel C) or higher probability of ranking all true features correctly (Figure 1 Pane B). To the best of my knowledge, this is the first simulation study that compared both unregularized and regularized Gaussian models with regularized Cox regression ($\lambda = \lambda_{min}$ or $\lambda_{1se}$), both in terms of feature selection accuracy and the accuracy of ranking effect sizes, under the assumptions that: 1) Time-to-event values are generated from multivariate Cox proportional hazards models, with different correlation levels between features, different sample sizes, different censoring rates and different parameter values for baseline risk function. 2) Not all true features are available for feature selection and effect size estimation. 3) Censored time occurs randomly and follows Uniform distributions. If one believes that we should assume unmeasured key factors exist outside of any given dataset from clinical trials or biomarker discovery studies, I think this simulation design provides some level of justification for the following proposed changes to survival analysis practice:

1. Instead of fitting univariate Cox regression separately for each measured feature, it might work better if we apply the Cox proportional hazards model with elastic net regularization ($\lambda=\lambda_{1se}$) to screen independent Gaussian features in clinical trials, even if only 10 features are measured and only 50 participants reported events of interest such as death or disease progression (Figure 1 Panel A).

2. For trials with modest number of events (50 to 500), it might work better if we fit the Gaussian regression using log-transformed survival time with two covariates (i.e., the event indicator plus one feature) to screen highly correlated Gaussian features, instead of using Cox proportional hazards models with elastic net regularization (Figure 1 Panel C). For trials with large number of events (>500), a better strategy might be fitting the multivariate Gaussian regression with elastic net regularization ($\lambda$=0.05) to screen highly correlated Gaussian features, even when the number of measured features is as small as 10.

3. Instead of performing both the univariate analysis and the multivariate analysis via Cox proportional hazards models without regularization, it might work better if we apply the Cox proportional hazards models with elastic net regularization ($\lambda=\lambda_{1se}$) to rank the effect sizes of highly correlated Gaussian features in clinical trials, even when only 10 features are measured and only 50 participants reported events of interest (Figure 1 Panel D)

In addition to my belief that unknown key factors exist outside of any dataset generated by clinical trials or biomarker discovery studies, another reason why I think it's imperative to share findings of this work is that this study compared models' performance over a wide range of small to modest event sizes, as well as various shapes of baseline risk functions. While the importance of studying asymptotic properties is well appreciated in methodology research of survival models, there might not be enough efforts devoted to studying model behavior given small/modest sample sizes. Considering why the Fisher's exact test and the Chi-squared test are recommended for testing independence using frequency tables with different cell/sample sizes and how many phase II clinical trials and phase III trials of rare diseases end with <500 events in survival analysis each year, I believe it would be very impactful even if a survival analysis idea could improve our current ability of analyzing small/modest datasets just a little bit. Figure 1 Panel C of this work at least provides a set of realistic scenarios where nonconventional survival models (Gaussian regression) should be recommended for feature selection tasks with respect for time-to-event outcomes if the total number of observed events is small. Moreover, depending on the specific goal of a study, models' performance in ranking effect sizes might be more important than its selection accuracy measured by (sensitivity + specificity)/2, especially for biomarker discovery studies with independent features and small sample sizes. But when features are highly correlated, (sensitivity + specificity)/2 could be more important if the correlation structure between features is invariant to interventions targeting any of these correlated features.

On the other hand, evidence provided by one simulation study is always very limited. Other interesting questions remain. For example, when stratified Cox regression are used in clinical trials, should we recommend fitting regularized Cox separately for different strata? How would different data preprocessing steps, such as normalizing or dichotomizing continuous covariates, impact feature selection accuracy and effect size estimation when sample size is small and not all factors are measured? Could Gaussian models outperform Cox regression when key features are multinomial? Even if all covariates are Gaussian, it's still unclear whether the recommendations proposed in this work could be further improved by better tuned regularization parameters or using tree-based method such as random forest [33-35]. Could certain tree-based methods outperform regression models when the true Cox model contains a large number of interaction terms between Gaussian covariates while many of these Gaussian variables are not measured? In order to make method recommendations for screening time-varying features or features with time-varying coefficients for small datasets, more simulation studies also need to be carefully designed with the assumption that many key features can't be measured.

# REFERENCES

1. Cliff, Edward R. Scheffer, and Ghulam Rehman Mohyuddin. "Overall survival as a primary end point in multiple myeloma trials." Nature Reviews Clinical Oncology (2022): 1-2.

2. Yang, Shenmiao, Neil E. Kay, Min Shi, Curtis A. Hanson, and Robert Peter Gale. "Is unmeasurable residual disease (uMRD) the best surrogate endpoint for clinical trials, regulatory approvals and therapy decisions in chronic lymphocytic leukaemia (CLL)?." Leukemia (2022): 1-5.

3. Parish, Austin J., Ioana Alina Cristea, Ewoud Schuit, and John PA Ioannidis. "2,109 randomized oncology trials map continuous, meager improvements in progression-free and overall survival over 50 years." Journal of Clinical Epidemiology 150 (2022): 106-115.

4. Li, Jianpei, Jianhua Lin, Yaoling Luo, Miaohuan Kuang, and Yijun Liu. "Multivariate analysis of prognostic biomarkers in surgically treated endometrial cancer." PLoS one 10, no. 6 (2015): e0130640.

5. Cox, D. R. "Regression Models and Life-Tables." Journal of the Royal Statistical Society. Series B (Methodological) 34, no. 2 (1972): 187–220. http://www.jstor.org/stable/2985181.

6. Peixoto, Carolina, Marta Martins, Luís Costa, and Susana Vinga. "Kidney Cancer Biomarker Selection Using Regularized Survival Models." Cells 11, no. 15 (2022): 2311.

7. Powles, Thomas, Srikala S. Sridhar, Yohann Loriot, Joaquim Bellmunt, Xinmeng Jasmine Mu, Keith A. Ching, Jie Pu et al. "Avelumab maintenance in advanced urothelial carcinoma: biomarker analysis of the phase 3 JAVELIN Bladder 100 trial." Nature medicine 27, no. 12 (2021): 2200-2211.

8. Oltra, Sara S., Maria Peña-Chilet, Victoria Vidal-Tomas, Kirsty Flower, María Teresa Martinez, Elisa Alonso, Octavio Burgues, Ana Lluch, James M. Flanagan, and Gloria Ribas. "Methylation deregulation of miRNA promoters identifies miR124-2 as a survival biomarker in Breast Cancer in very young women." Scientific reports 8, no. 1 (2018): 1-12.

9. Tibshirani, Robert. "The lasso method for variable selection in the Cox model." Statistics in medicine 16, no. 4 (1997): 385-395.

10. Friedman, Jerome, Trevor Hastie, and Rob Tibshirani. "Regularization paths for generalized linear models via coordinate descent." Journal of statistical software 33, no. 1 (2010): 1.

11. Simon, Noah, Jerome Friedman, Trevor Hastie, and Rob Tibshirani. "Regularization paths for Cox's proportional hazards model via coordinate descent." Journal of statistical software 39, no. 5 (2011): 1.

12. Tibshirani, Robert, Jacob Bien, Jerome Friedman, Trevor Hastie, Noah Simon, Jonathan Taylor, and Ryan J. Tibshirani. "Strong rules for discarding predictors in lasso-type problems." Journal of the Royal Statistical Society: Series B (Statistical Methodology) 74, no. 2 (2012): 245-266.

13. Ternès, Nils, Federico Rotolo, and Stefan Michiels. "Empirical extensions of the lasso penalty to reduce the false discovery rate in high-dimensional Cox regression models." Statistics in medicine 35, no. 15 (2016): 2561-2573.

14. Hastie, Trevor, Robert Tibshirani, and Ryan Tibshirani. "Best subset, forward stepwise or lasso? Analysis and recommendations based on extensive comparisons." Statistical Science 35, no. 4 (2020): 579-592.
15. Mazumder, Rahul. "Discussion of "best subset, forward stepwise or lasso? analysis and recommendations based on extensive comparisons"." Statistical Science 35, no. 4 (2020).
16. Hastie, Trevor, Robert Tibshirani, and Ryan J. Tibshirani. "Rejoinder: Best Subset, Forward Stepwise or Lasso? Analysis and Recommendations Based on Extensive Comparisons." Statistical Science 35, no. 4 (2020): 625-626.
17. Lu, Rong, Danxin Wang, Min Wang, and Grzegorz A. Rempala. "Estimation of Sobol's sensitivity indices under generalized linear models." Communications in Statistics-Theory and Methods 47, no. 21 (2018): 5163-5195.
18. Knief, Ulrich, and Wolfgang Forstmeier. "Violating the normality assumption may be the lesser of two evils." Behavior Research Methods 53, no. 6 (2021): 2576-2590.
19. Riepenhausen, Antje, Ilya M. Veer, Carolin Wackerhagen, Zala C. Reppmann, Göran Köber, José Luis Ayuso-Mateos, Sophie A. Bögemann et al. "Coping with COVID: risk and resilience factors for mental health in a German representative panel study." Psychological medicine (2022): 1-11.
20. Chen, Jie, Kees de Hoogh, John Gulliver, Barbara Hoffmann, Ole Hertel, Matthias Ketzel, Mariska Bauwelinck et al. "A comparison of linear regression, regularization, and machine learning algorithms to develop Europe-wide spatial models of fine particles and nitrogen dioxide." Environment international 130 (2019): 104934.

21. Papili Gao, Nan, SM Minhaz Ud-Dean, Olivier Gandrillon, and Rudiyanto Gunawan. "SINCERITIES: inferring gene regulatory networks from time-stamped single cell transcriptional expression profiles." Bioinformatics 34, no. 2 (2018): 258-266.

22. Austin, Peter C., Jiming Fang, and Douglas S. Lee. "Using fractional polynomials and restricted cubic splines to model non-proportional hazards or time-varying covariate effects in the Cox regression model." Statistics in Medicine 41, no. 3 (2022): 612-624.

23. Zhang, Chao, Xiaoyong Li, Feng Li, Gugong Li, Guoqiang Niu, Hongyu Chen, Guang-Guo Ying, and Mingzhi Huang. "Accurate prediction and further dissection of neonicotinoid elimination in the water treatment by CTS@ AgBC using multihead attention-based convolutional neural network combined with the time-dependent Cox regression model." Journal of Hazardous Materials 423 (2022): 127029.

24. Spreafico, Marta, Francesca Ieva, and Marta Fiocco. "Modelling time-varying covariates effect on survival via functional data analysis: application to the MRC BO06 trial in osteosarcoma." Statistical Methods & Applications (2022): 1-28.

25. Thackham, Mark, and Jun Ma. "On maximum likelihood estimation of competing risks using the cause-specific semi-parametric Cox model with time-varying covariates–An application to credit risk." Journal of the Operational Research Society 73, no. 1 (2022): 5-14.

26. Wu, Wenbo, Jeremy MG Taylor, Andrew F. Brouwer, Lingfeng Luo, Jian Kang, Hui Jiang, and Kevin He. "Scalable proximal methods for cause-specific hazard modeling with time-varying coefficients." Lifetime Data Analysis 28, no. 2 (2022): 194-218.
27. Zhao, Jimmy L., Karim Fizazi, Fred Saad, Kim N. Chi, Mary-Ellen Taplin, Cora N. Sternberg, Andrew J. Armstrong, Johann S. de Bono, William T. Duggan, and Howard I. Scher. "The Effect of Corticosteroids on Prostate Cancer Outcome Following Treatment with Enzalutamide: A Multivariate Analysis of the Phase III AFFIRM Trial." Clinical Cancer Research 28, no. 5 (2022): 860-869.
28. Haruki, Koichiro, Tomohiko Taniai, Mitsuru Yanagaki, Kenei Furukawa, Masashi Tsunematsu, Shinji Onda, Yoshihiro Shirai, Michinori Matsumoto, Norimitsu Okui, and Toru Ikegami. "Sustained systemic inflammatory response predicts survival in patients with hepatocellular carcinoma after hepatic resection." Annals of Surgical Oncology (2022): 1-10.
29. Fu, Ningzhen, Kai Qin, Jingfeng Li, Jiabin Jin, Yu Jiang, Xiaxing Deng, and Baiyong Shen. "Who could complete and benefit from the adjuvant chemotherapy regarding pancreatic ductal adenocarcinoma? A multivariate-adjusted analysis at the pre-adjuvant chemotherapy timing." Cancer Medicine (2022).
30. Qing, Liangliang, Qingchao Li, Yongjin Yang, Wenbo Xu, and Zhilong Dong. "A prognosis marker MUC1 correlates with metabolism and drug resistance in bladder cancer: A bioinformatics research." BMC urology 22, no. 1 (2022): 1-13.

31. Pasqualetti, Francesco, Celeste Giampietro, Nicola Montemurro, Noemi Giannini, Giovanni Gadducci, Paola Orlandi, Eleonora Natali et al. "Old and New Systemic Immune-Inflammation Indexes Are Associated with Overall Survival of Glioblastoma Patients Treated with Radio-Chemotherapy." Genes 13, no. 6 (2022): 1054.
32. Lu, Rong. "Should univariate Cox regression be used for feature selection with respect to time-to-event outcomes?." arXiv preprint arXiv:2208.09689 (2022).
33. Breiman, Leo. "Random forests." Machine learning 45, no. 1 (2001): 5-32.
34. Hothorn, Torsten, Peter Bühlmann, Sandrine Dudoit, Annette Molinaro, and Mark J. Van Der Laan. "Survival ensembles." Biostatistics 7, no. 3 (2006): 355-373.
35. Ishwaran, Hemant, Udaya B. Kogalur, Eugene H. Blackstone, and Michael S. Lauer. "Random survival forests." The annals of applied statistics 2, no. 3 (2008): 841-860.

**SUPPLEMENTARY TABLES AND FIGURES**

- **Supplementary Table 1:** Summary of sensitivity, specificity and probability of correctly ranking all true features
- **Supplementary Figure 1**: Line plots of feature selection sensitivity and specificity
- **Example simulation script:** sim_7_09172022_n500.R
- **Clarification script**: sim_6_10082022_n1000v2.R (with baseline risk function parameters < 2)
- **Supplementary Figure 2**: 3D Scatter Plot of R function *coxph()* Sensitivity vs. Cox Regression Baseline Risk Function Parameters

<u>Supplementary Figure 1</u>: Line Plots of Sensitivity and Specificity

**(A)**
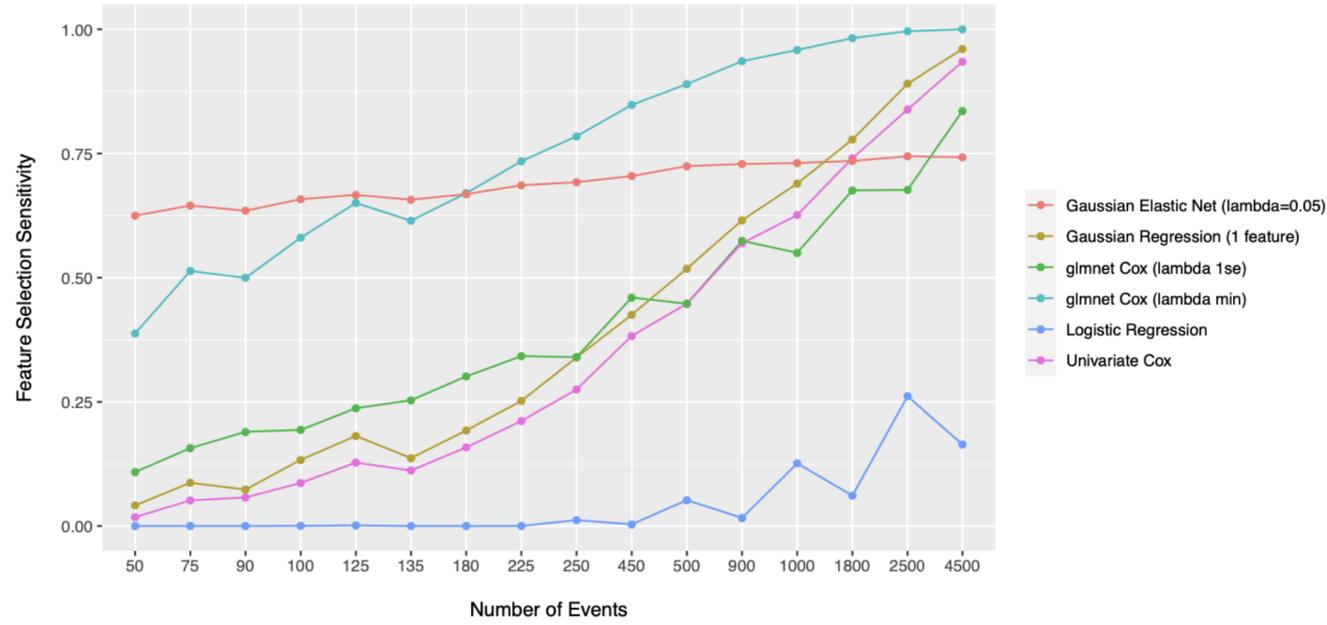

**(B)**
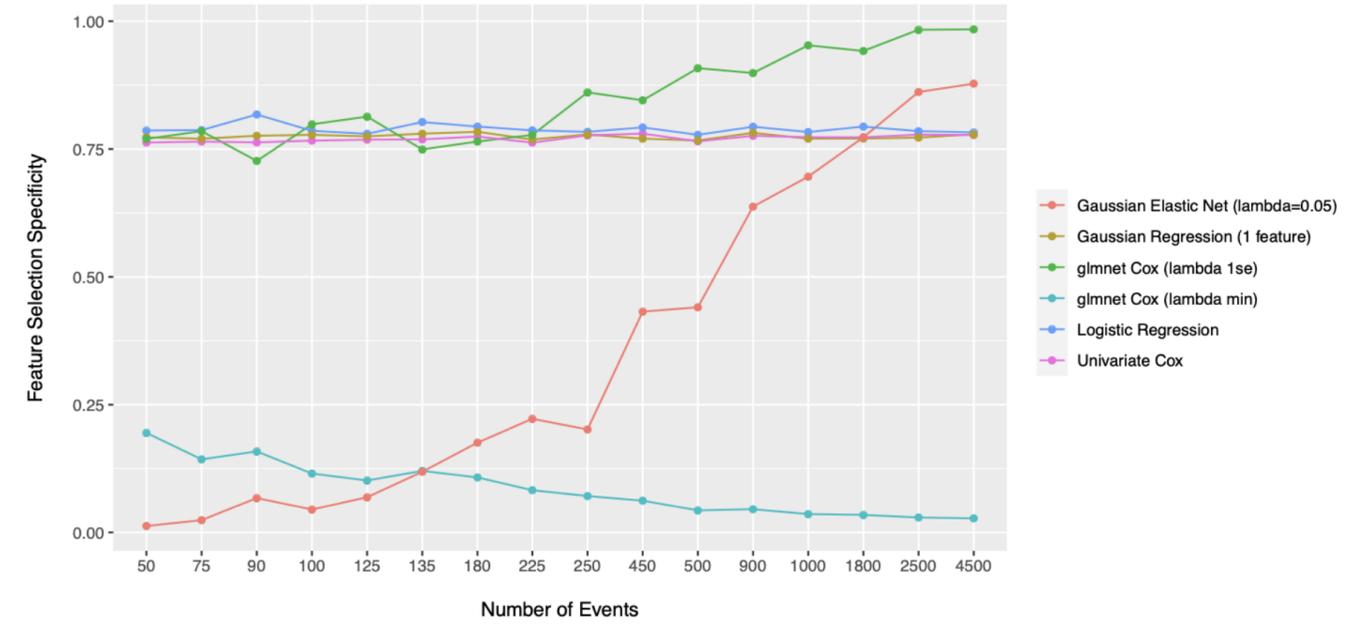

**(C)**
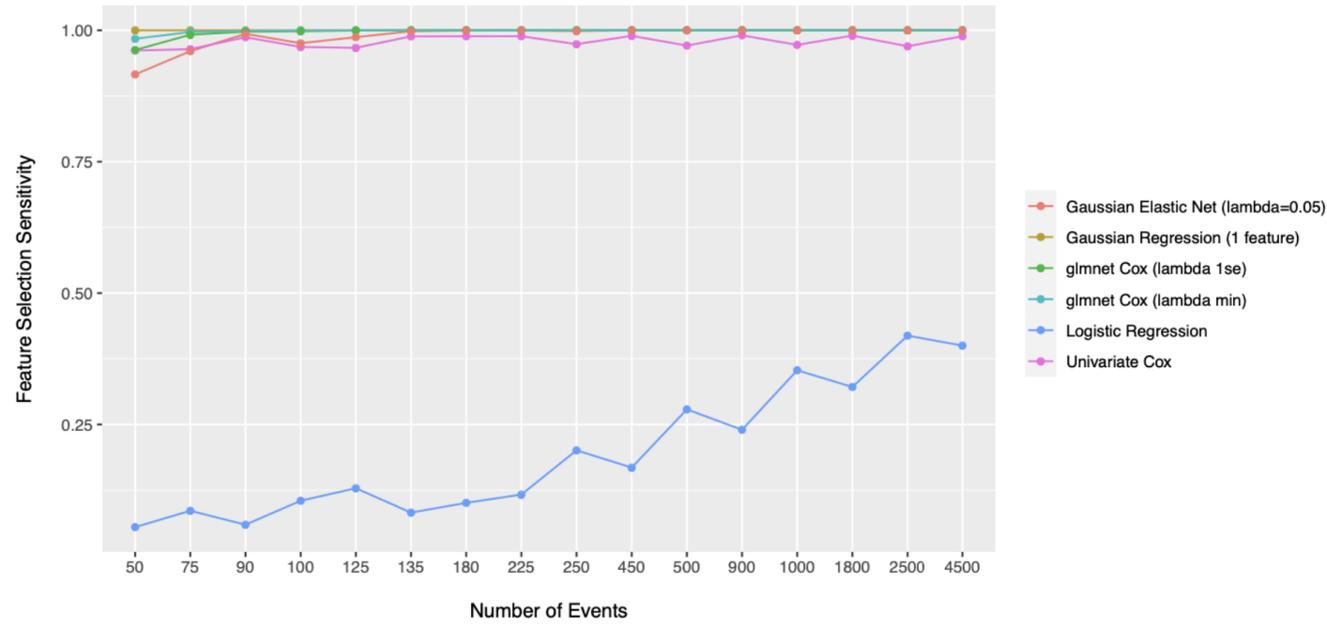

**(D)**
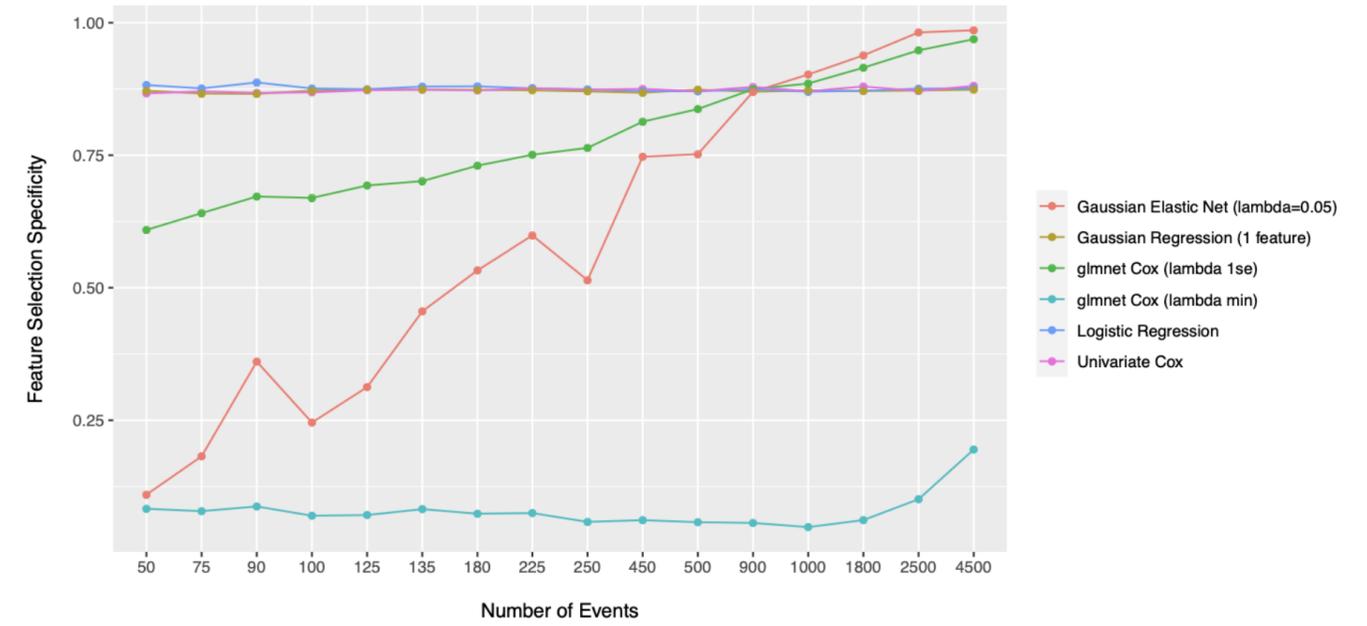

**Supplementary Figure 2**: 3D Scatter Plot of R function *coxph()* Sensitivity vs. Baseline Risk Function Parameters

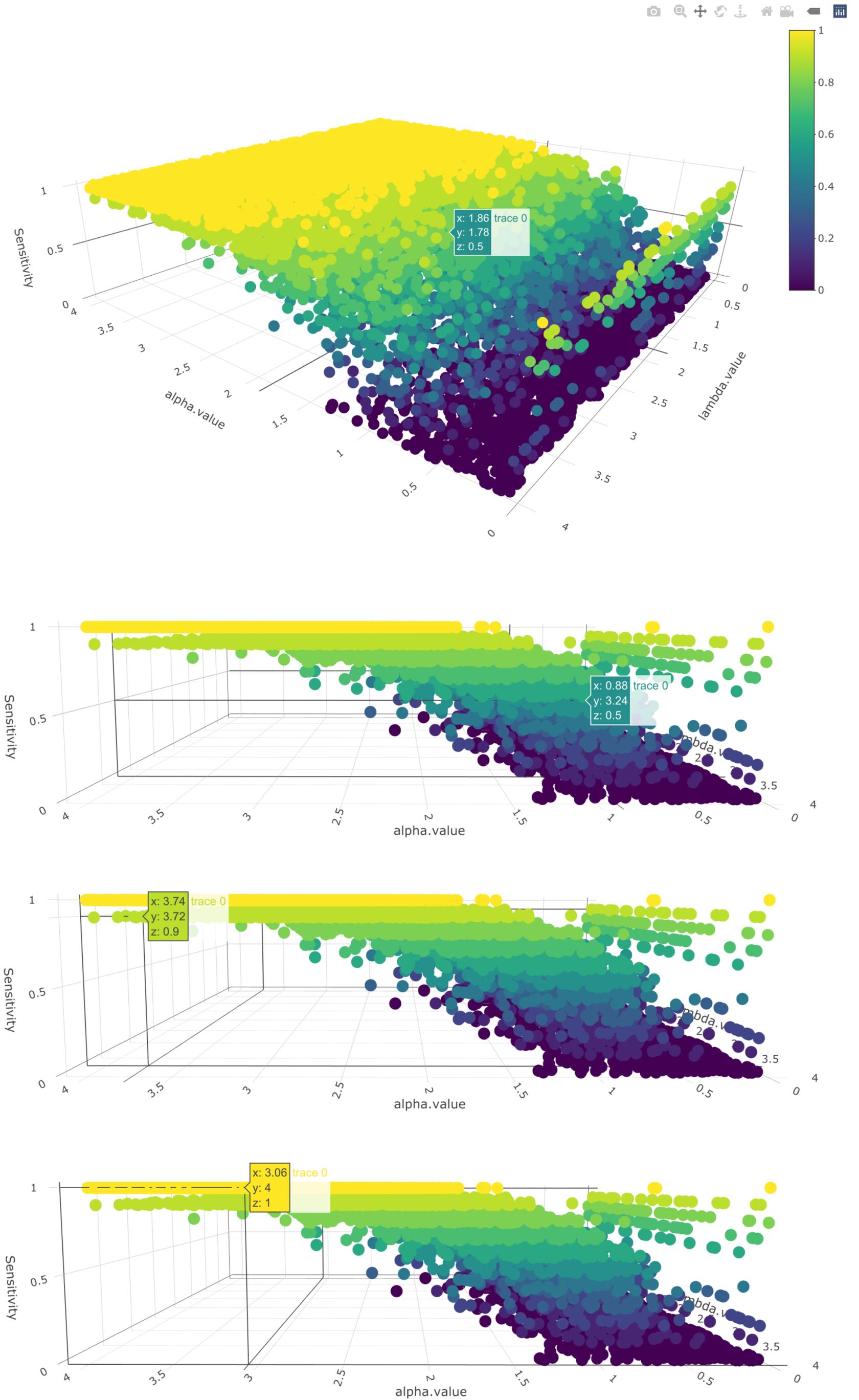